\documentclass[10pt]{iopart}
\begin{document}
\title[The one-loop elastic coefficients$\,\dots$]{The one-loop elastic 
coefficients for the Helfrich membrane in higher dimensions}
\author{J~A~Santiago${}^1$ and A~Zamora${}^2$}
\address{${}^1$ Centro de Investigaci\'on Avanzada en Ingenier\'\i a 
Industrial. Universidad Aut\'onoma del Estado de Hidalgo, Pachuca 42090,
M\'exico.}
\address{${}^2$ Instituto de Ciencias Nucleares, Universidad Nacional 
Aut\'onoma de M\'exico, Apartado Postal 70-543, M\'exico DF 04510, 
M\'exico.}
\eads{\mailto{sgarciaj@uaeh.reduaeh.mx}, \mailto{zamora@nuclecu.unam.mx}}

\begin{abstract}
Using a covariant geometric approach we obtain the effective bending
couplings for a $2$-dimensional rigid membrane embedded into a
$(2+D)$-dimensional Euclidean space. The Hamiltonian for the membrane
has three terms: The first one is quadratic in its mean extrinsic
curvature. The second one is proportional to its Gaussian curvature, 
and the last one is proportional to its area. The results we obtain 
are in agreement with those finding that thermal fluctuations soften
the $2$-dimensional membrane embedded into a $3$-dimensional Euclidean
space.
\end{abstract}

\submitto{\JPA}

\pacs{87.16.Dg, 46.70.Hg}


\section{Introduction}
\label{I}
Because of their chemical structure, when one adds amphiphilic
molecules in an aqueous substance they spontaneously form
membrane structures known as micelles of various geometrical
shapes (spherical or cylindrical in particular). The relevance of 
these structures, in part, is that they can be regarded as simplified 
models of biological molecules and, that they can actually be created 
artificially in the laboratory for its use in cosmetics and detergents 
\cite{Lipowsky}. Despite the diversity of these amphiphilic membranes, 
they can be modeled by a single geometric Hamiltonian that captures 
their relevant physical properties as the surface tension $\mu$ and 
rigidity $\kappa$. For the last, a term that penalizes the energetic cost 
of giving curvature to the membrane has been suggested \cite{Hel, Can}.
Surprisingly, geometric Hamiltonians of this sort of surfaces appear 
in several areas of physics ranging from string theory and cosmology 
to condensed matter \cite{Chaikin, Vilenkin, Polchinski, Polyakov}. 

In equilibrium, the membrane configurations satisfy the Euler-Lagrange 
equations which are obtained by extremizing the Hamiltonian respect 
to the membrane orthogonal deformations. These equations are essentially 
geometric in its own nature. The evaluation of the second variation of 
the Hamiltonian is necessary to study the stability of equilibrium 
solutions and to understand the effect of fluctuations on the
shape of the membrane. If one wishes to understand, for instance,
the transitions between different shapes that the membrane can have,
one has to study the effect of small fluctuations on the membrane
shape. A problem in statistical physics, in this context, is to study
the membrane elasticity under thermal fluctuations; i.e. to investigate
the effect of such fluctuations on the bending constants; the membrane
rigidity in particular \cite{nelson,safran,peliti}. This problem has
already been considered by a number of authors 
\cite{helfrich,peliti2,foster,kleinert,borelli}. The conclusion they
reach is that thermal fluctuations soften the membrane. For a 
quasi-two-dimensional membrane there exists a non-trivial fixed point $T_c\,$:
A membrane with temperature lower than $T_c$ is very rigid at long length 
scales, but one with temperature greater than $T_c$ has very low rigidity
(becoming zero for length scales beyond a characteristic length 
called the persistence length).
For exactly two-dimensional membranes, which is the critical dimension
of the model, it can be interpreted that the membrane disappears as
a consequence of fluctuations. This is similar to what occurs in the
$O(N)$ non-linear sigma model where, in two dimensions, fluctuations
destroy any attempt of ordering.
Recently, however, it has been suggested that thermal fluctuations 
rather stiffen the membrane \cite{pinnow,Nishiyama} instead of 
softening it. The interpretation of this result is not clear yet. So, 
because of this, we have consider it useful to investigate the subject 
using an alternative approach. 

In principle, there are two different approaches which allow one to
study the effect of thermal fluctuations on the membrane. One of them;
the one we use here, consists of a perturbative expansion for
finite $D$ codimension with the inverse of the rigidity $T/\kappa$ 
($T$ the temperature) being the perturbative parameter. Our results
indicate that thermal fluctuations lower the value of the rigidity,
so the model falls into the two-dimensional membrane scenario 
described above. The other approach consists of a non-perturbative
approximation in $D$ within the $D\to\infty$ limit \cite{DavidD}.
As pointed out in Ref. \cite{HKleinert}, and by analogy with 
the $O(N)$ nonlinear sigma model, the non-perturbative $D\to\infty$ limit 
approximation must somehow regain the properties of a two-dimensional 
membrane in a three-dimensional Euclidean space within the one-loop 
approximation. In particular, the one-loop result for the membrane
rigidity of codimension $D$ is exact within the $D\to\infty$ limit
\cite{DavidD}. This rises the question whether this is still valid
to order two-loops. The approach we propose in this article may,
amongst other things, give an answer to this question or it can
help to understand, for instance, the effect of quantum fluctuations
in geometric models from string theory as those proposing the
rigidity term as fundamental for their description \cite{fazio}.

To be explicit, the work in this article consists in obtaining,
in the one-loop order, the elastic bendings for a $2$-dimensional 
membrane embedded into a $(2+D)$-dimensional Euclidean space.
Though it is possible to generalize for higher dimensional
membranes, we consider this the most relevant case. The calculation
of the elastic bendings, to this order, starts by determining 
the operator of fluctuations associated with small deformations in 
the shape of the membrane. A common approach for that is to consider 
the Monge representation where the Hamiltonian is expanded in terms of 
the membrane height, $z=h(x,y)$. Beyond the Gaussian model, perturbation 
theory yields autointeraction terms extremely difficult to deal with. 
It is important to realise, however, that carrying out a perturbative 
expansion can be advantageous indeed as it accounts for the crumpling
transition \cite{peliti}. There is the disadvantage, though, of loosing 
covariance in the expressions appearing at intermediate calculations.

For this investigation we use an alternative formalism, capable of dealing 
with geometric models such as the Helfrich model for fluid membranes, 
which does preserve covariance in all expressions. This formalism has 
been extensively used before by Capovilla and Guven \cite{capovilla}, 
and is found particularly powerful in simplifying long calculations. 
As an example, within this formalism the Euler-Lagrange equations can 
be obtained in a few lines from a geometric Hamiltonian. Moreover, 
this fully covariant approach allows one to solve several technically 
difficult problems such as the description of lipid membranes with an 
edge and to tackle the problem of membrane adhesion 
\cite{capovilla2,capovilla3}. In addition to those, it has recently 
been used to investigate thermal fluctuations of fluid membranes in 
higher dimensions. As a result, the operator of fluctuations 
${\cal L}^{ij}$ for a $N$-dimensional membrane embedded into an 
Euclidean space of dimension $(N+D)$ has been found in a systematic 
manner \cite{capoguvsan}.

The Hamiltonian associated with the model under study is written as a 
sum of three terms: The first one is quadratic in the mean extrinsic 
curvature, $K^{i}$. The second one is proportional to the scalar 
curvature of the membrane $\cal R$, and the last one proportional to 
its area $A$,
\begin{equation}
{\cal H}_0={\kappa\over 2}\int\rmd A\,\,  K^iK_i
           +\beta\int\rmd A\, {\cal R}+\mu\int\rmd A\,. 
\label{elfrich}
\end{equation}
Given that $[K]=L^{-1}$, any other terms of higher order in the curvature
are irrelevant for two-dimensional membranes in the long-length regime.
This ${\cal H}_0$ above is a generalized model of a lipid membrane with 
non-zero rigidity and surface tension which has also been studied in the
string theory context \cite{kleinert1, polyakov}. The model is invariant
under translations and rotations in space.
The partition function $Z$ is constructed as a functional integral over 
all possible surface configurations, ${\bf X}$, of the form
\begin{equation}
Z=\int {\cal D}{\bf X}\,\, \exp\left(-{\cal H}_0/T\,\right)
\label{Partition}\,, 
\end{equation}
where $T$ is the temperature parameter. In the one-loop order, we expand
up to second order the exponent in equation (\ref{Partition}) around a 
configuration extremizing ${\cal H}_0$. Notice that, as discussed in
reference \cite{Lubensky}, even in the one-loop order the integral measure
contributes with a non-linear factor (a sort of Liouville term) that has to be
taken into account. Leaving this for later consideration and after 
computing a Gaussian integral, the one-loop effective Hamiltonian may 
be formally written as
\begin{equation}
{\cal H}_{e\!f\!\!f}={\cal H}_0+{T\over 2}{\rm Tr}\,\log\,{\cal L}^{ij}\,,
\label{Hef}
\end{equation} 
with ${\cal L}^{ij}$ the operator for small fluctuations and {\rm Tr}
representing the functional trace. Subsequently, we obtain the one-loop
effective bending coefficients of the model and present the results
below. As it will be shown, they are in agreement with those obtained
for the case of a two-dimensional membrane embedded into a 
three-dimensional Euclidean space. 

The organization of this paper is as follows. As the fundamental tool
used throughout this work is essentially geometry, we describe in 
section \ref{GoES} the geometry of embedded surfaces. In section 
\ref{DitG} we study deformations in the quantities present in the 
Hamiltonian. Further, the Euler-Lagrange equations of the problem are
also written here. In section \ref{SoD} we present the operator of 
fluctuations for the Hamiltonian and in section \ref{KeR} we obtain 
the one-loop effective parameters. Finally, our results are summarized 
in section \ref{S}.

\section{Geometry and Deformations}
\label{GoES}
An orientable $N$-dimensional surface $\Sigma$, embedded into the 
$(N+D)$-dimensional Euclidean space $R^{N+D}$, is locally specified by 
the embedding functions,
\begin{equation}
{\bf x}={\bf X}(\xi^a)\,,
\end{equation}
where ${\bf X}=(X^{1},\cdots ,X^{N+D})$ are $(N+D)$ functions of $N$
variables each. The metric $\gamma_{ab}$, induced on $\Sigma$, is
defined as 
\begin{equation}
\gamma_{ab}:={\bf e}_a\cdot{\bf e}_b\,,
\end{equation}
with ${\bf e}_a=\partial_a{\bf X}$ ($\partial_a=\partial/\partial\xi^a$)
and we are using the inner product in $R^{N+D}$. This metric
$\gamma_{ab}$ determines the intrinsic geometry of $\Sigma$.

The Riemann curvature ${\cal R}^a_{\,\,\,{bcd}}$ is defined as
\begin{equation}
(\nabla_a\nabla_b-\nabla_b\nabla_a)V^c={\cal R}^c_{\,\,\,{dab}}V^d\,,
\end{equation}
with $\nabla_a$ being the covariant derivative associated to 
$\gamma_{ab}$. The Ricci tensor is defined by 
${\cal R}^{ab}:={\cal R}^c_{\,\,\,{acb}}$ and the scalar curvature 
given by ${\cal R}:=\gamma_{ab}{\cal R}^{ab}$.

Let us now examine the extrinsic geometry of $\Sigma$. The $D$
unit vectors ${\bf n}_{i}(\xi)$, normal to $\Sigma$ in $R^{N+D}$,
are defined implicitly by
\begin{equation}
{\bf e}_{a}\cdot {\bf n}^{i}=0\,,  \label{eq:dn1}
\end{equation}
which, after normalization, satisfy 
\begin{equation}
{\bf n}^{i}\cdot {\bf n}^{j}=\delta ^{ij}\,.  \label{eq:dn2}
\end{equation}
Notice that these equations determine the normals ${\bf n}^{i}$ up to
an arbitrary rotation. Therefore, the definition is invariant under
local rotations,
\begin{equation}
{\bf n}^i \to O^i{}_j (\xi^a) {\bf n}^j \,.
\end{equation}
The set of vectors $\{{\bf n}^{i},{\bf e}_{a}\}$ is precisely an adapted
base on the surface. Their gradients along the surface are written in 
terms of this base as the Gauss-Weingarten equations
\begin{eqnarray}
\partial _{a}{\bf e}_{b} &=&\gamma _{ab}{}^{c}{\bf e}_{c}-K_{ab}^{i}%
{\bf n}_{i}\,,  \label{eq:gw1} \\
\partial _{a}{\bf n}_{i} &=&K_{ab\,i}\gamma ^{bc}{\bf e}_{c}+\omega
_{ij\,a}{\bf n}^{j}\,.  \label{eq:gw2}
\end{eqnarray}
The extrinsic curvatures of $\Sigma$ are given by the $D$ 
symmetric rank-two surface tensors, $K_{ab\,i}:=-{\bf n}_{i}\cdot 
\partial _{a}{\bf e}_{b}=K_{ba\,i}\,$. We also define the trace with
respect to the induced metric,
\begin{equation}
K_{i}:=\gamma ^{ab}K_{ab\,i}\,.
\end{equation}

Both the intrinsic and extrinsic geometry of $\Sigma$, which are 
determined by $\gamma_{ab}$, $K_{ab\,i}$ and $\omega_{ij\,a}$, can no
longer be specified independently. They are related by integrability
conditions that correspond to generalized Gauss-Codazzi and
Codazzi-Mainardi equations, and also by equations that fix the
curvature $\Omega_{ijab}$ associated with the $SO(D)$ connection
$\omega_{ija}$ in terms of $K_{ab}^{i}$:
\begin{eqnarray}
{\cal R}_{abcd}-K_{ac\,i}K_{bd}^{i}+K_{ad\,i}K_{bc}^{i} &=&0\,,
\label{eq:gauss1} \\
\tilde{\nabla}_{a}K_{bc}^{i}-\tilde{\nabla}_{b}K_{ac}^{i} &=&0\,,
\label{eq:cm1} \\
\Omega _{ij\,ab}+K_{ac\,i}K^{c}{}_{b\,j}-K_{ac\,j}K^{c}{}_{b\,i} &=&0\,.
\label{ricci}
\end{eqnarray}
In these expressions $\tilde{\nabla}_{a}$ is the $SO(D)$ covariant
derivative defined as
\begin{equation}
\tilde{\nabla}_{a}\phi _{i}=\nabla _{a}\phi _{i}
     -\omega _{ij\,a}\phi ^{j}\,.
\label{gcd}
\end{equation}

\section{Deformations in the Geometry}
\label{DitG}

\subsection{The Metric and the Extrinsic Curvature}
\label{TMatEC}

We proceed as in the hypersurface case and project the deformation onto
its normal and tangent components in the form \cite{capovilla5} 
\begin{equation}
\delta {\bf X}=\Phi^{i}{\bf n}_{i}+\Phi^{a}{\bf e}_{a}\,.
\end{equation}
By now we will assume that the surface has no boundary so that the 
tangent deformation can be associated to a reparametrization. The only
physical deformations are those normal to the surface. Let us then 
examine the deformation in the adapted base 
$\{{\bf e}_{a},{\bf n}^{i}\}$ under these normal deformations. Therefore,
the relationships
\begin{eqnarray}
{\delta}_{\perp}{\bf e}_{a}&=&(\tilde{\nabla}_{a}\Phi^{i}){\bf n}_{i}
+\Phi ^{i}K_{ia}{}^{c}{\bf e}_{c}\,,  \label{A}\\
{\delta}_{\perp }{\bf n}^{i}&=&-(\tilde{\nabla}^{a}\Phi ^{i}){\bf e}
_{a}\,,
\end{eqnarray}
hold. In those we have used the covariant derivative $\tilde{\nabla}$ 
defined in equation (\ref{gcd}). By using equation (\ref{A}), it is straightforward to 
show that the deformation in the induced metric is given by
\begin{eqnarray}
{\delta}_{\perp }\gamma _{ab} &=&( {\delta}_{\perp }{\bf e}_{a})\cdot 
{\bf e}_{b}+{\bf e}_{a}\cdot ( {\delta}_{\perp }{\bf e}_{b})  \nonumber\\
&=&2K_{iab}\Phi ^{i}\,.  \label{a}
\end{eqnarray}
From this, we get the first order deformation of the infinitesimal 
area element
\begin{equation}
\delta_{\perp}dA=dA K^i\Phi_i\,.  \label{p}
\end{equation}
Let us now turn to the first order deformation of the extrinsic curvature
along the ${\it i}{\rm th}$ normal vector. For that we find
\begin{eqnarray}
{\delta}_{\perp }K_{ab}^{i} &=&-( {\delta}_{\perp }{\bf n}%
^{i})\cdot\partial _{a}{\bf e}_{b}-{\bf n}^{i
}\cdot\partial _{a}
{\delta}_{\perp }%
{\bf e}_{b}  \nonumber \\
&=&-{\nabla}_{a}{\nabla}_{b}\Phi^{i}
+K_{ac}{}^{i}K^{c}{}_{bj}\Phi ^{j}\,.
\end{eqnarray}
As a result, the deformation in the mean extrinsic curvature is
\begin{equation}
{\delta}_{\perp }K^{i}=-\tilde{\Delta}\Phi
^{i}-K_{ab}^{i}K^{ab}{}_{j}\Phi ^{j}\,.
\end{equation}
From these expressions it is straightforward to find the corresponding 
Euler-Lagrange equations of the model (${\cal E}^{j}=0$), which are 
given by \cite{capoguvsan}
\begin{equation}
{\cal E}^{j}=-\kappa \tilde{\Delta }K^{j}+{\kappa\over 2} \left(
2R^{ij}-K^{i}K^{j}\right) K_{i}-2\beta {\cal G}_{ab}K^{abj}+\mu K^{j}\,.
\end{equation}

\subsection{Second Order Deformations}
\label{SoD}

To second order we express the result in terms of the deformations
already in hand. Therefore, given the Hamiltonian in equation (\ref{elfrich}),
we find that the operator of fluctuations can be written in the
compact form
\begin{equation}
{\cal L}_{ij}= \kappa\tilde\Delta^2_{ij} + A^{ab\,ik} 
(\tilde\nabla_a \tilde\nabla_b)_{kj} + B^{a}_{ik}(\tilde \nabla_a)_{kj}
+ C_{ij}\, ,\label{ele}
\end{equation}
with the coefficient of the Laplacian term given by $(K^{2}=K_{i}K^{i})$
\begin{eqnarray}
A^{ab\, ij} &=&{\kappa\over 2}\left( 2K^{i}K^{j}-
K^{2}\delta ^{ij}-4R^{ij}\right)\gamma^{ab}-\mu\delta^{ij}\gamma^{ab} 
+2\kappa K_\ell K^{\ell ab}\delta^{ij}\nonumber\\
&+&2\beta\left( {\cal G}^{abji}+{\cal G}^{baij}
+{\cal G}^{ab}\delta^{ij}\right)\,;
\label{aij} 
\end{eqnarray}
where the definitions
\begin{eqnarray}
{\cal G}_{ab}{}^{ij} &=&R_{ab}{}^{ij}-{\frac{\gamma _{ab}}{2}}R^{ij}\,,\\
R_{ab}{}^{ij} &=&K_{ab}^{i}K^{j}-K_{a}{}^{ci}K_{cb}{}^{j} \,,
\end{eqnarray}
have been used. Both coefficients ${\cal B}$ and ${\cal C}$ contain 
$K\nabla K$ and $K^4$ terms respectively \cite{capoguvsan}; which
will modify, in any case, terms of higher order in the membrane 
curvature. These, however, have not been taken into account in the 
Hamiltonian because they are irrelevant for long-length scales for
two-dimensional membranes.

\section{The Heat Kernel Regularization}
\label{KeR}

The heat kernel provides a covariant method for calculating the short 
distance behaviour of the effective Hamiltonian 
\cite{dewitt, barbin,birreldavis}. One examines the heat equation 
associated with the Laplacian on the surface. Its kernel satisfies
\begin{equation}
\left({\partial\over \partial s}+ 
\tilde \Delta \right){\cal K}_{ij}(\xi, \xi', s)=0\,,
\end{equation}
subject to the initial condition
\begin{equation}
{\cal K}_{ij}(\xi , \xi' , 0)={\delta (\xi ,\xi' )\over \sqrt\gamma}
\delta_{ij}\,.
\end{equation}
The solution to this equation can be written in the form
\begin{equation}
{\cal K}^{ij}(\xi , \xi' , s)={1\over (4\pi s )^{N/2}}
\,\,\exp\left( -{\sigma (\xi , \xi' )\over 2s} \right) 
{\cal W}^{ij}(\xi , \xi', s)\,{\cal D}^{1/2}(\xi ,\xi')\,,
\end{equation}
with $\sigma (\xi , \xi')$ being the world function (equal to half the 
squared geodesic distance between the points $\xi$ and $\xi'$) and
${\cal D}(\xi , \xi')$ the Van Vleck-Morette determinant. In the $s\to 0$
limit, the exponential normalized by $1/ (4\pi s)^{N/2}$ reduces to
the delta function. We thus have that $\cal W$ is a regular function of
its arguments which may be expanded in power series as
\begin{equation}
{\cal W}^{ij}(\xi , \xi', s)=\sum_{l=0}^{\infty}s^l 
a_l^{ij}(\xi , \xi')\,. 
\end{equation}
In the coincidence limit, $\xi'\to\xi$, the first three 
coefficients $a_l^{ij}(\xi )$ are given by
\begin{eqnarray}
&&a_0^{ij}= \delta^{ij}\,, \quad  a_1^{ij}(\xi )=
{{\cal R}\over 6}\,\delta^{ij}\,,\\ 
&& a_2^{ij}(\xi )= \left({1\over 180}
\left({\cal R}_{abcd}^2-{\cal R}_{ab}^2\right) 
+ {{\cal R}^2\over 72} + {\Delta {\cal R}\over 30}\right)\,\delta^{ij}+ 
{1\over 2} \Omega{}^{i}{}_k{}^{ab}\Omega{}^{kj}{}_{ab}\,.  
\end{eqnarray}
Henceforth we will absorb the determinant ${\cal D}^{1/2}(\xi ,\xi')$ 
into the definition of the $a_l's$. Despite the undifferentiated 
coincidence limits are the same, one may need to be careful with 
coincidence limits of derivatives. The corresponding Green function
associated with the Laplacian can then be expanded as
\begin{equation}
\tilde\Delta^{-1}(\xi, \xi')^{ij} =
-\int {\rmd s \over (4\pi s)^{N/2}} 
\,\,\exp\left( -{\sigma (\xi , \xi' )\over 2s} \right)
\sum_{l=0}^\infty
s^l a_l^{ij}(\xi,\xi' )\,.
\end{equation}
Notice that the trace over fluctuations in the range 
$k_{\rm min}, k_{\rm max}$ is implemented by integrating $s$ over the 
range, $k_{\rm max}^{-2}, k_{\rm min}^{-2}$. The heat kernel can be 
exploited to regularize the trace in the problem. For 
instance, it can be written in the coincidence limit
\begin{equation}
{\rm Tr}\,\log \left( -\tilde\Delta \right) =-\int\rmd A\,\, 
\int {\rmd s\over s(4\pi s)^{N/2}}
\sum_{l=0}^\infty s^l a_l^{ii}(\xi )\label{trln}\,,
\end{equation}
and the following expression of great use below:
\begin{equation}
\Delta^{-2}(\xi,\xi')^{ij}=  
\int {\rmd s \quad s\over (4\pi s)^{N/2}}
\,\,\exp\left(-{ \sigma\over 2s }\right) \sum_{l=0}^\infty
s^l a_l(\xi,\xi' )^{ij}\label{D2}\,.
\end{equation}
All this allows us to calculate the second term in equation (\ref{Hef}).
So, taking the logarithm of equation (\ref{ele}) and expanding we find
\begin{eqnarray}
{\rm Tr}\,\log\left( {\cal L}^{ij}\right)
&\simeq&{\rm Tr}\,\log\left(\kappa\tilde\Delta^2_{ij}\right)
+ {1\over\kappa}\,{\rm Tr}\, A^{ab\,ik} 
\left(\tilde\nabla_a \tilde\nabla_b\right)_{k\ell} 
(\tilde\Delta^{-2})^{\ell j}\nonumber\\
&+&{1\over\kappa}\,{\rm Tr}\, B^{aik}\tilde\nabla_{ak\ell}
(\tilde\Delta^{-2})^{\ell j}+{1\over \kappa}{\rm Tr}\,
C_{ik}(\tilde\Delta^{-2})^{k j}\,. \label{cuatro}
\end{eqnarray}
As mentioned in section \ref{I}, the contribution of the measure is 
still to be taken into account. To order one loop, that is given by
minus the first term in equation (\ref{cuatro}), so its contribution
cancels out the first term \cite{Lubensky}.

By using the expression for the operator of fluctuations in equation 
(\ref{aij}), it is easy to show that the last two terms are at 
least of the order $K^4$, and therefore can just be dropped out.
Therefore, the only contribution for the renormalization comes from 
the second term. By taking two derivatives of equation (\ref{D2}), we get 
that in the coincidence limit, $\xi'\to\xi$,
\begin{eqnarray}
{\rm Tr}\, A^{ab} \tilde\nabla_a\tilde\nabla_b   \tilde\Delta^{-2} &=&
\sum_{l=0}^\infty
\int {\rmd s\, s^{l+1}\over (4\pi s)^{N/2}}
\int\rmd AA^{ab} \tilde\nabla_a\tilde\nabla_b 
\left\{  a_l \,\,\exp \left( -{\sigma\over 2s }\right)\right\}\nonumber\\ 
&=& \sum_{l=0}^\infty\int {\rmd s s^l\over (4\pi s )^{N/2}}\int\rmd A  \, 
A^{ab\, ij}\left(s\nabla_a\nabla_b-{\gamma_{ab}\over 2}\right)
a_l^{ij}\,.\label{trAab}
\end{eqnarray}
The second line follows because, in the coincidence limit, only the term 
appearing from differentiating $\sigma$ twice survives:
\begin{equation}
\tilde\nabla_a\tilde\nabla_b \left(  a_l 
\,\,\exp \left( -{\sigma\over 2s }\right)
\right)= \tilde\nabla_a\tilde\nabla_b a_l -{\gamma_{ab}\over 2s} a_l\,.
\end{equation}
From equations (\ref{trAab}) and (\ref{aij}) we obtain the one loop 
effective bendings for the Helfrich membrane for arbitrary codimension.
For $N=2$ (i.e. for a two-dimensional membrane) only the terms with 
$l=0$ must be taken into account to renormalize any ultraviolet
singularity, $s\to 0$. Moreover, the Van Vleck-Morette determinant does
not contribute at all. In addition, from equation (\ref{aij}) we see that
\begin{equation}
A^a{}_{ajj}={\kappa\over 2} \left(2(2+D)\right)K^2-2\mu D-4
\kappa{\cal R}\,,
\end{equation}\label{aajj}
for a two-dimensional membrane embedded into a $(2+D)$-dimensional
Euclidean space. Using this result into equation (\ref{trAab}) we
find the effective bending coupling which describes the crumpling
transition, see e.g. reference \cite{nelson}. For the rigidity we have  
\begin{equation}
\kappa_{e\!f\!\!f}=\kappa -T{2+D\over 4\pi}\log {L\over a}\,,
\end{equation}
where $L$ is the linear size of the membrane and $a$ is a microscopic
cutoff. If the codimension is set to $D=1$, this equation reproduces
precisely the result already found in references 
\cite{peliti2, foster, kleinert, borelli}. In the string theory context,
and using a different approach, this result was found in references
\cite{kleinert1, polyakov}. For the surface tension we obtain
\begin{equation}
\mu_{e\!f\!\!f}=\mu +T{D\over 4\pi}{\mu\over\kappa}
\log {L\over a}\label{tension}\,.
\end{equation}
A result which coincides with that found in reference \cite{kleinert} (also 
for arbitrary codimension) and with that from reference \cite{David} 
(obtained for $D=1$). This method provides a way to identify the 
$\beta_{e\!f\!\!f}$ parameter also for arbitrary $D$. For a two-dimensional
membrane we get
\begin{equation}
\beta_{e\!f\!\!f}=\beta +{2T\over 4\pi}\log {L\over a}\,,
\end{equation}
which is also in agreement with the result found in reference \cite{kleinert}
for $D=1$ and, as expected, the renormalization of this $\beta$ 
coefficient does not depend on $D$. Notice that if the 
contribution of the measure is not taken into account, the term 
${\rm Tr}\,\log (-\tilde\Delta)$ gives a spurious one-loop contribution to
the renormalization of both $\mu_{e\!f\!\!f}$ and $\beta_{e\!f\!\!f}$. Moreover,
for a two-dimensional membrane the Gaussian term does not give a 
contribution as it should because in this case the term is a topological
invariant.

\section{Summary}
\label{S}

In this paper we have used an extended covariant geometric approach 
recently developed to obtain the one loop effective bending couplings of
a geometric Hamiltonian associated with a $2$-dimensional membrane 
embedded into a $(2+D)$-dimensional space. Our result is in agreement 
with that appearing in the literature for a two-dimensional membrane
embedded into a three-dimensional Euclidean space. Work in the two-loop
calculation is now underway.

\ack
One of us, JAS, is grateful to Jemal Guven for many discussions and to
Prof. H. Kleinert for kind hospitality at the Institute f\"ur 
Theoretische Physik, Freie Universit\"at Berlin. AZ acknowledges
hospitality of the Instituto de Ciencias Nucleares, UNAM. This work was
supported by the PROMEP program PTC in Mexico.

\end{document}